\documentclass[superscriptaddress,pra,amsmath,twocolumn,amssymb]{revtex4}

\usepackage{dcolumn}
\usepackage{amsfonts}
\usepackage{amssymb}
\usepackage{amsmath}
\usepackage{amsthm}
\usepackage{latexsym}
\usepackage{epsfig}
\usepackage{color}
\usepackage{lscape}
\usepackage{multirow}
\usepackage{braket}
\usepackage{bbm}
\usepackage{mathrsfs}
\usepackage{fancyhdr}
\usepackage{lastpage}

\newcommand{\red}{\color{black}}

\newcommand{\be}{\begin{equation}}
\newcommand{\ee}{\end{equation}}
\newcommand{\ba}{\begin{array}}
\newcommand{\ea}{\end{array}}
\newcommand{\bea}{\begin{eqnarray}}
\newcommand{\eea}{\end{eqnarray}}

\begin{document}

\title{Testing randomness with photons}
\author{Jonathan C. F. Matthews}
\affiliation{Centre for Quantum Photonics, H. H. Wills Physics Laboratory
and Department of Electrical and Electronic Engineering, University
of Bristol, Merchant Venturers Building, Woodland Road,
Bristol BS8 1UB, UK.}
\author{Rebecca Whittaker}
\affiliation{Centre for Quantum Photonics, H. H. Wills Physics Laboratory
and Department of Electrical and Electronic Engineering, University
of Bristol, Merchant Venturers Building, Woodland Road,
Bristol BS8 1UB, UK.}
\author{Jeremy L. O'Brien}
\affiliation{Centre for Quantum Photonics, H. H. Wills Physics Laboratory
and Department of Electrical and Electronic Engineering, University
of Bristol, Merchant Venturers Building, Woodland Road,
Bristol BS8 1UB, UK.}
\author{Peter S. Turner}
\email{peter.turner@bristol.ac.uk}
\affiliation{Department of Physics, Graduate School of Science, University of Tokyo, 7-3-1 Hongo, Bunkyo-ku, Tokyo, Japan 113-0033\footnote{Now at the School of Physics, H. H. Wills Physics Laboratory, Tyndall Avenue, University of Bristol, Bristol BS8 1TL, UK.}}
\date{\today}

\begin{abstract}
Generating and characterising randomness is fundamentally important in both classical and quantum information science.
Here we report the experimental demonstration of ensembles of pseudorandom optical processes comprising what are known as $t$-designs.
We show that in practical scenarios, certain finite ensembles of two-mode transformations---1- and 2-designs---are indistinguishable from truly random operations for 1- and 2-photon quantum interference, but they fail to mimic randomness for 2- and 3-photon cases, respectively.
This provides a novel optical test of pseudorandomness.
We make use of the fact that $t$-photon behaviour is governed by degree-$2t$ polynomials in the parameters of the optical process to experimentally verify the ensembles' behaviour for \emph{complete} bases of polynomials.
This ensures that outputs will be uniform for arbitrary configurations, satisfying the strict definition of pseudorandomness implicit in the mathematical definition.
\end{abstract}

\maketitle

Randomness underpins science and technology, from simulating {\red complex} systems and modelling error through to {\red probabilistic} computation and information security. 
In quantum mechanics, randomness is a fundamental feature,
but {\red additional} randomness {\red can be} a powerful resource when {\red purposely introduced into} quantum {\red protocols; examples include} 
quantum communication~\cite{ab-prsa-465-2537}, {\red quantum algorithms}~\cite{ra-alg-55-490}, quantum data hiding~\cite{ha-cmp-250-371}, benchmarking unknown quantum processes~\cite{em-sci-302-2098,em-sci-317-1893,ep-arxiv-1308.2928}, and the boson sampling conjecture~\cite{aaroson-acm}.
However, truly random quantum operators are inefficiently realisable both in principle and in practice. 
Here we report the realisation and complete characterisation of {\red two} examples of 
\emph{pseudorandom} photonic quantum operator ensembles that provide a practical alternative: {\red so-called} \emph{$t$-designs} that simulate statistical properties of truly random operators {\red using} fewer resources~\cite{gr-jmp-48-052104,da-pra-012304}. 
We make use of the fact that $t$-photon behaviour is governed by degree-$2t$ polynomials in optical process parameters to experimentally verify the ensembles' behaviour.
We realise a 1-design and a 2-design, 
and show that 1- and 2-photon quantum interference~\cite{ho-prl-59-2044,ma-apb-60-s111} is sufficient for their complete verification.
We further show that 2- and 3-photon interference{\red , respectively, can be used to test} the limits of their pseudorandom properties.
We apply these ideas in a realistic scenario to characterise randomness when standard quantum process tomography fails, demonstrating a new application of photons.

Fair and unbiased random quantum processes---transformations from one {\red pure} quantum state to another---requires sampling uniformly from the continuously infinite group of unitary operations on a system.
This group is equipped with a unique invariant (Haar) measure, which defines the uniform distribution over the set of all unitaries.
However, such ``true'' randomness is inefficiently realisable in {\red practice}, due to poor scaling of the number of random parameters with the size of the system.
Fortunately, randomly sampling uniformly from a restricted subensemble of unitary operations can be done efficiently and still exhibits some of the desired statistical properties of truly random processes.
Such pseudorandom operations are therefore sought after for applications in quantum protocols.
This notion of pseudorandomness is captured well by \textit{unitary $t$-designs}.

Roughly speaking, unitary $t$-designs are subensembles of quantum operations that, given $t$ copies of a system, are statistically indistinguishable from a uniformly distributed superensemble.
Equivalently, they are subensembles that have the same $t$-th order moments as the uniform Haar ensemble, and thus they can be used to simulate some statistical properties of truly random quantum operations with fewer resources.
These statistical moments are given by polynomials in the parameters of the quantum operators in question, and a strict definition in terms of such polynomials is made below.

\begin{figure}[]
\centering
\includegraphics[width=
\columnwidth]{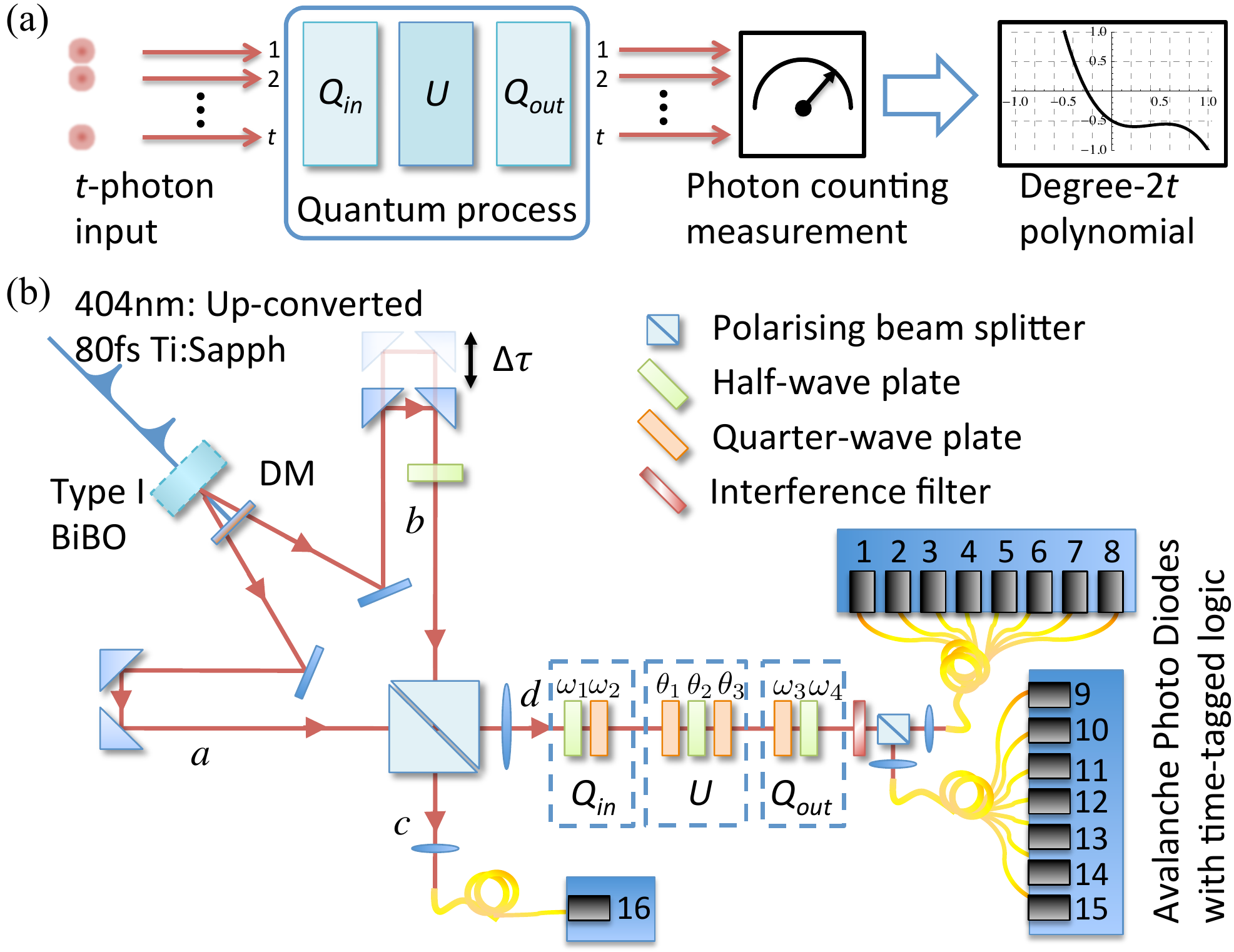}
\vspace{-0.5cm}
\caption{\footnotesize{\textbf{Sampling a polynomial with a quantum photonic process.}
(a) Probabilities for $t$-photon interference are described by degree-{$2t$} polynomials.
By choosing different input $Q_\mathrm{in}$ and output $Q_\mathrm{out}$ configurations, it is possible to access different polynomials.
(b) Our experimental setup was used to sample complete sets of independent degree-$2$ and degree-$4$ polynomials in the elements of the SU(2) rotation $U$ using 1- and 2-photon interference by post-selecting 1- and 2-photon states from spontaneous parametric downconversion (SPDC, see Appendix).
Three-photon experiments were also performed to sample degree-6 polynomials.
Ideally identical photons generated by non-linear spontaneous parametric downconversion in paths $a$ and $b$ are combined onto $d$ and two polarisation modes using a half-wave plate and a polarising beam splitter cube.
The quantum process $T=Q_\mathrm{out}UQ_\mathrm{in}$ is {\red r}ealised using a collection of quarter- and half-wave plates {\red (dashed boxes)}.
}}
\label{setup}
\end{figure}

Here we are concerned specifically with the superensemble of all unitary polarisation rotations of an optical channel, which can be expressed as matrices parameterised by the real variables $\{x_1,y_1,x_2,y_2\}$ such that
\begin{eqnarray}
U=\left(
\begin{array}{c c}
x_1 + i y_1 & x_2 + i y_2\\
-x_2 + i y_2 & x_1 - i y_1
\end{array}
\right),
\label{SU2matrix}
\end{eqnarray}
with the constraint~\footnote{This parameterisation actually gives the SU$(2)$ subgroup of the unitary group, but the distinction---a global phase---will not concern us here.} $x_1^2+y_1^2+x_2^2+y_2^2 = 1$.
The probability distributions that govern the measurement outcomes of any multiphoton interference experiment using such operations are polynomials in the matrix elements of $U$. The degree of these polynomials is dictated by the number of photons (Fig.~\ref{setup}(a)).
For example, the transition probability of one photon from input 2 to output 2 is  $|U_{2,2}|^2 = x_1^2+y_1^2$, which is a degree-2 polynomial in the parameters.
Note that this exactly agrees with how the intensity of classically modelled light is distributed as it passes through $U$.
Two-photon non-classical interference experiments are governed by degree-4 polynomials, including the probability to detect coincidentally one photon in each output of $U$ in a Hong-Ou-Mandel experiment~\cite{ho-prl-59-2044}, which is given by $|U_{1,1}U_{2,2}+U_{1,2}U_{2,1}|^2$~\cite{ma-apb-60-s111}.
In general, $t$-photon interference is modelled by a degree-$2t$ polynomial in the matrix elements of the unitary process~\cite{li-njp-7-155}, and for which there is no classical description without either some reduction in the visibility of features amongst correlated detection patterns~\cite{br-prl-102-253904} or an overhead of resources~\cite{ke-pra-81-023834}.

A unitary $t$-design is defined in terms of such polynomials~\cite{da-pra-012304}.
Explicitly, a finite set $\mathfrak{D_t}$ containing $K$ unitary operators, viewed as an ensemble with uniform distribution $1/K$, is defined to be a $t$-design if every degree-$2t$ polynomial~\footnote{Elsewhere $t$-designs have been defined such that a distinction is made between the degrees of conjugate variables, but here we use only the real variables $\{x_1,y_1,x_2,y_2\}$.} in the matrix elements of $U$, $f_{2t}(U)$, has the same average over $\mathfrak{D_t}$ as it does when averaged over the uniform ensemble of \emph{all} unitaries,
\begin{eqnarray}
\mathbb{E}_\mathfrak{D_t}[f_{2t}] \! = \! \! \!\! \sum_{U \in \mathfrak{D}_t} \frac{1}{K} f_{2t}(U) = \!\! \int \! \textrm{d} U  \, f_{2t}\left(U\right) \! = \!  \mathbb{E}_\mathrm{Haar}[f_{2t}].
\label{tdesigndefintion}
\end{eqnarray}
Uniformity in the continuous case is defined by the normalised unitary Haar measure d$U$, and there are several methods for computing the integral over the unitary group, (e.g.~ref.~\onlinecite{co-cmp-264-773}). 
Note that a $t$-design is by definition also a $(t-1)$-design, hence
experiments with $t$ or fewer photons sampled over a $t$-design are statistically indistinguishable from the same experiments with operations sampled from the Haar distribution. 
The transition probabilities of multiphoton interference can therefore be used to verify the realisation of a $t$-design.
{\red Additionally, because} Eq.~(\ref{tdesigndefintion}) does not in general hold for $(>\!\!2t)$-degree polynomials, $(>\!\!t)$-photon interference can also be used to test pseudorandomness as an alternative to process tomography.

We realise unitary 1- and 2-designs in two dimensions using linear optics.
The ensemble of operators we use to realise a 1-design is
\begin{eqnarray}
\mathfrak{D}_1 = \left\{I, i X, -i Y, i Z \right\},
\label{Design1}
\end{eqnarray}
the uniformly distributed Pauli ensemble~\footnote{The phases were chosen to give a representation of the unit quaternions $1 \leftrightarrow I$, $i \leftrightarrow iX$, $j \leftrightarrow -iY$ and $k \leftrightarrow iZ$.}.
Intuitively, for the case of a Haar distributed random ensemble, we expect input laser light intensity (and single photons) to be distributed uniformly across the outputs---for SU$(2)$ rotations on average half of the laser intensity input into mode 1 will transmit to output 2, i.e. $\mathbb{E}_\mathrm{Haar}[|U_{1,1}|^2]=1/2$.
Sampling uniformly over $\mathfrak{D}_1$ would agree with this: for example, light input into 1 will be distributed equally between output 1 (due to $I$ and $Z$) and output 2 (due to $X$ and $Y$).
It is perhaps not so intuitive to see that this is true for arbitrary inputs due to the interference of complex amplitudes, but it is straightforward to verify that $\mathfrak{D}_1$ is a 1-design using Eq.~(\ref{tdesigndefintion}) by computing the average over a complete basis of polynomials of degree-2. 
For the four real-valued variables $x_1, y_1, x_2, y_2$ and the single constraint that $U$ is unitary, there are nine independent degree-2 basis monomials, which we list in the Appendix.

It is intuitive from a quantum optics perspective to see $\mathfrak{D}_1$ fails to behave like the Haar distributed random ensemble in two-photon interference experiments, and is therefore not a 2-design.
Consider a Hong-Ou-Mandel experiment where we estimate the probability for two indistinguishable photons input into ports 1 and 2 to anti-bunch at the two outputs \cite{ho-prl-59-2044}.
If $U$ corresponds to a 50:50 beamsplitter then there is ideally zero probability to detect one photon at each output.
Therefore, when sampling over the entire Haar ensemble the average probability for the photons to anti-bunch must be strictly less than 1.
Averaging over $\mathfrak{D}_1$, however, will always yield an anti-bunching probability of exactly 1, hence Eq.~(\ref{tdesigndefintion}) does not hold.

{\red T}he twelve {\red uniformly distributed} operators~\cite{gr-jmp-48-052104}
\begin{eqnarray}
\mathfrak{D}_2 &=& \{I, i X, -i Y, i Z, (I \pm i X \pm {i Y} \pm {i Z})/2\}\label{Design2}
\end{eqnarray}
form a 2-design, which is straightforward to verify by showing Eq.~(\ref{tdesigndefintion}) is satisfied
with respect to a complete set of 25 independent degree-$4$ monomials{\red .}
We implement each element of $\mathfrak{D}_2$ and $\mathfrak{D}_1$ using a combination of two quarter-wave plates (QWP) and one half-wave plate (HWP) as shown in the box labeled $U$ in Fig.~\ref{setup} (b).  Explicit  
wave plate settings for each element are provided in the Appendix.

In order to verify that a physically realised ensemble of operators is a $t$-design, we need to 
experimentally access complete sets of polynomials and average them over 
the elements of the ensemble being tested.
We have found that for SU(2), complete sets of degree-2 and degree-4 polynomials are physically accessible using the arrangement in Figure~\ref{setup}(b).
We realise the polynomials by estimating different photon scattering distributions that
correspond to different input configurations $Q_\mathrm{in}$ and output or detection configurations $Q_\mathrm{out}$.
Thus, for each polynomial setting $(Q_\mathrm{in}, Q_\mathrm{out})$ and design element $U$, we have a unitary transformation $T := Q_\mathrm{out} U Q_\mathrm{in}$, realised as a polarisation interferometer.
We inject states of 1, 2, and 3 indistinguishable photons, collecting the number statistics at the output, and from the data we compute normalised photon scattering distributions, corrected for characterised circuit and detection efficiencies.

\begin{figure}[t!]
\includegraphics[width=1\columnwidth]{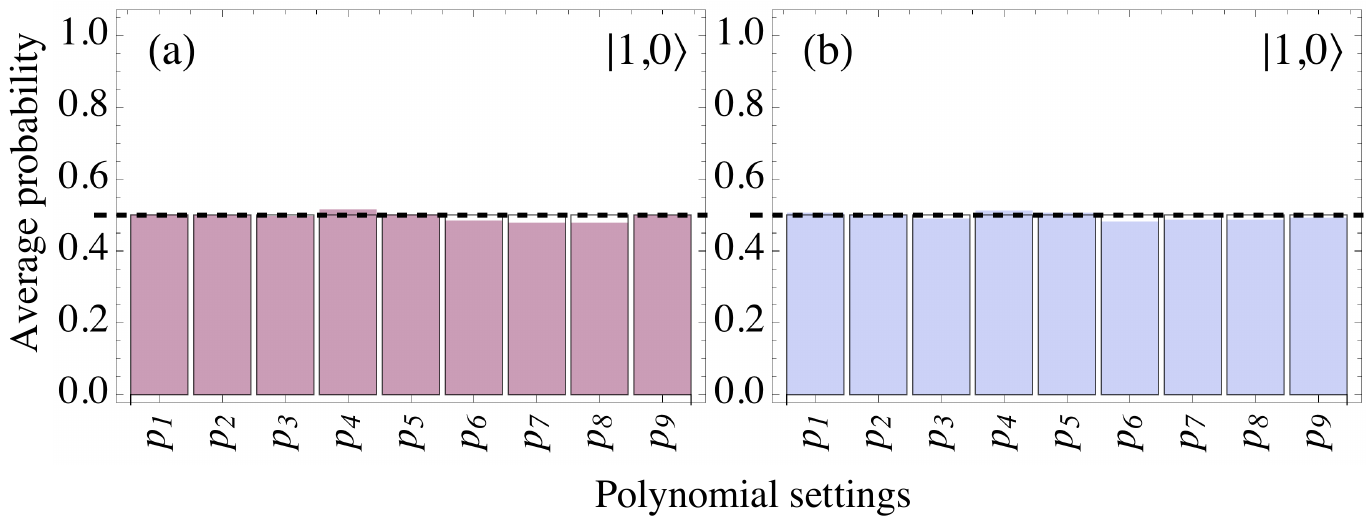}
\vspace{-0.7cm}
\caption{
\footnotesize{\textbf{Verifying pseudorandomness with 1 photon.} $|T_{1,1}|^2$ averaged over (a) $\mathfrak{D}_1$, and (b) $\mathfrak{D}_2$, showing both agree well with the Haar distribution. Here (and in Fig.~\ref{2photonAverage},\ref{3photonAverage}) both the experimentally extracted distributions (solid colour) and the corresponding ideal theoretical values for $\mathfrak{D}_t$ (empty boxes) are plotted. The dashed line in each plot represents the ideal Haar value, which is always uniform for normalised probability distributions.  See the Appendix for a discussion of quantifying the uniformity.
}}
\label{1photonAverage}
\end{figure}

The wave plate angles that realise a complete set of 1-photon polynomials were found by numerically searching for linearly independent polynomials $|{T}_{1,1}|^2$.
Assuming the quadratic constraint, there are nine such polynomials, which we label $p_1,...,p_9$; for example
$p_1=x_1^2+{y_1}^2$.
Similarly, 25 physically accessible linearly independent polynomials of degree-$4$ of the form $|{T}_{1,1}{T}_{2,2} + {T}_{1,2} {T}_{2,1}|^2$ were also found and are labeled $q_1,...,q_{25}$, where for example
$q_1=x_1^4-2 x_1^2 x_2^2+2 x_1^2 y_1^2+x_2^4-2 x_2^2 y_1^2+y_1^4$.
We provide the wave plate settings for all polynomials accessed in our demonstrations in the Appendix.

Fig.~\ref{1photonAverage} shows 1-photon probability distributions $|T_{1,1}|^2$ extracted from the experiment, taken for the 9 independent degree-2 polynomial settings $p_1,..,p_9$ and averaged uniformly over the ensembles $\mathfrak{D}_1$ (Fig.~\ref{1photonAverage} (a)) and $\mathfrak{D}_2$ (Fig.~\ref{1photonAverage} (b)).
For each polynomial setting, the single photon detection events are collected for each element of $\mathfrak{D}_1$ and $\mathfrak{D}_2$, from which normalised probability distributions are computed for each outcome, taking into account characterised output channel loss and relative detection efficiency.
We then average each distribution over $\mathfrak{D}_1$ and $\mathfrak{D}_2$; both agree with the uniform {\red Haar average over all unitaries}.
Since the 9 polynomials tested are a complete set, this agreement verifies the two ensembles $\mathfrak{D}_1$ and $\mathfrak{D}_2$ are both at least unitary 1-designs.

\begin{figure}
\includegraphics[width=\columnwidth]{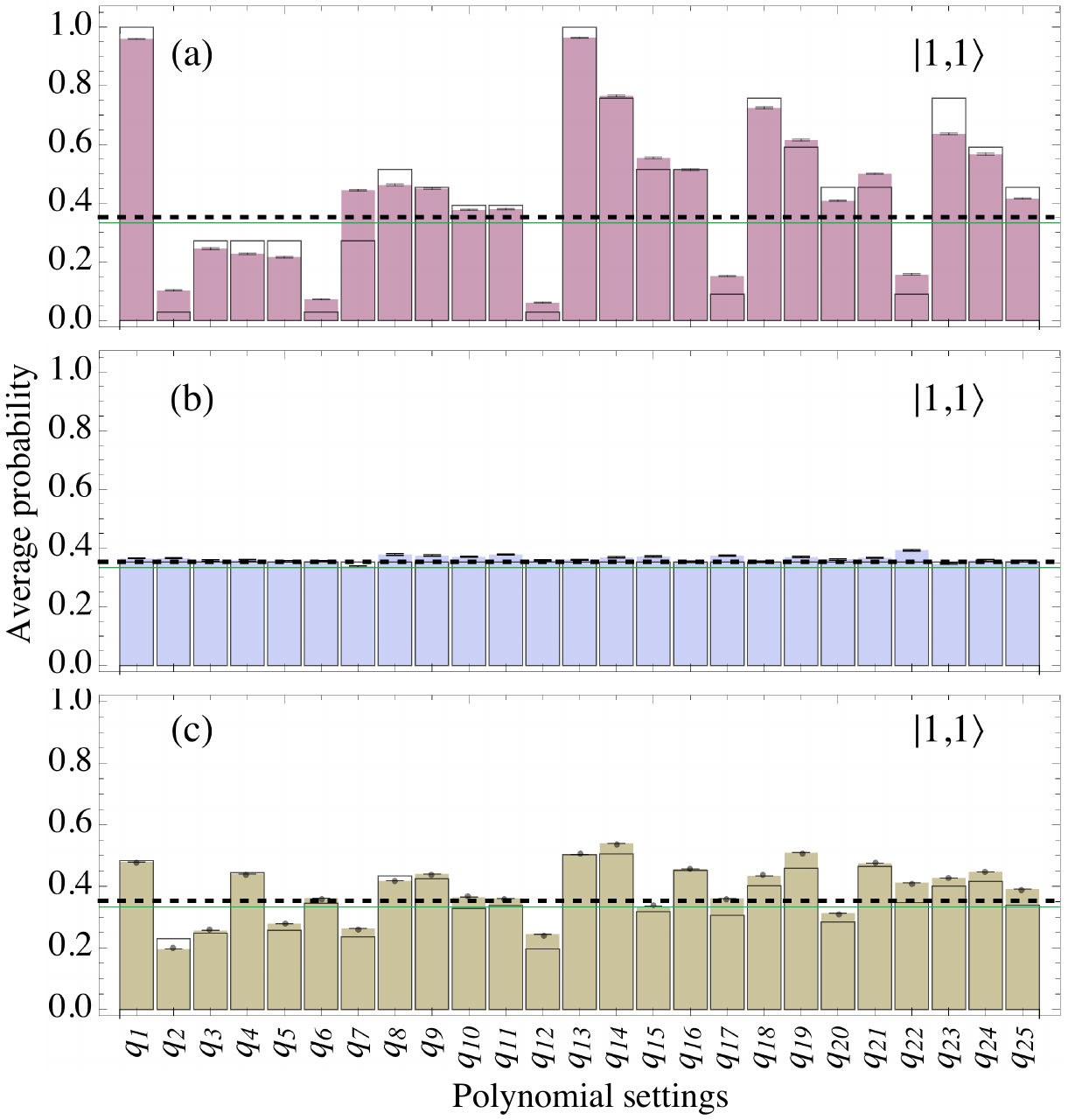}
\vspace{-0.7cm}
\caption{
\footnotesize{\textbf{Testing pseudorandomness with 2 photons.}
$|T_{1,1}T_{2,2}+T_{1,2}T_{2,1}|^2$ averaged over (a) $\mathfrak{D}_1$ and (b) $\mathfrak{D}_2$, showing that for two photons $\mathfrak{D}_2$ agrees well with the Haar distribution while $\mathfrak{D}_1$ does not. For comparison, (c) shows the averages over twelve unitary operators chosen randomly from the Haar distribution that do not form a 2-design.  Now photon distinguishability is an issue, altering the realised polynomials from their ideal values (green line).  This is characterised in the experiment and the polynomials are corrected accordingly (dashed line, see the Appendix for details).
}}
\label{2photonAverage}
\end{figure}

We see a clear difference in the behaviour of $\mathfrak{D}_1$ and $\mathfrak{D}_2$ when observing 2-photon interference statistics.
Fig.~\ref{2photonAverage} shows the 2-photon probability distributions $|T_{1,1}T_{2,2}+T_{1,2}T_{2,1}|^2$ extracted from experiment, taken for the 25 independent degree-2 polynomial settings $q_1,..,q_{25}$ and averaged uniformly over the ensembles $\mathfrak{D}_1$ (Fig.~\ref{2photonAverage} (a)) and $\mathfrak{D}_2$ (Fig.~\ref{2photonAverage} (b)).
The average of degree-4 polynomials over $\mathfrak{D}_1$ shows behaviour clearly distinct from averaging over the Haar measure (black dashed lines).
Together with the results of Fig.~\ref{1photonAverage} (a), this agreement verifies the ensemble $\mathfrak{D}_1$ is a 1-design only.
However, the uniformity of the degree-4 averages over $\mathfrak{D}_2$ (Fig.~\ref{1photonAverage} (b)) agrees closely with the average over the Haar measure.
Since $q_1,...,q_{25}$ is a complete set of degree-4 polynomials, this verifies the ensemble $\mathfrak{D}_2$ is at least a 2-design.
We also show in Fig.~\ref{2photonAverage} (c) 2-photon interference statistics averaged over a set of 12 matrices chosen randomly from the Haar distribution~\cite{Me-NAMS-54-5}.
The data demonstrate that in general an ensemble of twelve operations---the size of $\mathfrak{D}_2$---is not sufficient to simulate the Haar average, though of course a larger ensemble eventually will (see refs~\onlinecite{ha-cmp-264-773,br-arxiv-1208.0692} for related convergence results).  We expect that as $t$ increases, it will be harder to distinguish between $t$- and $(t+1)$-designs due to experimental noise.

\begin{figure}[t!]
\includegraphics[width=\columnwidth]{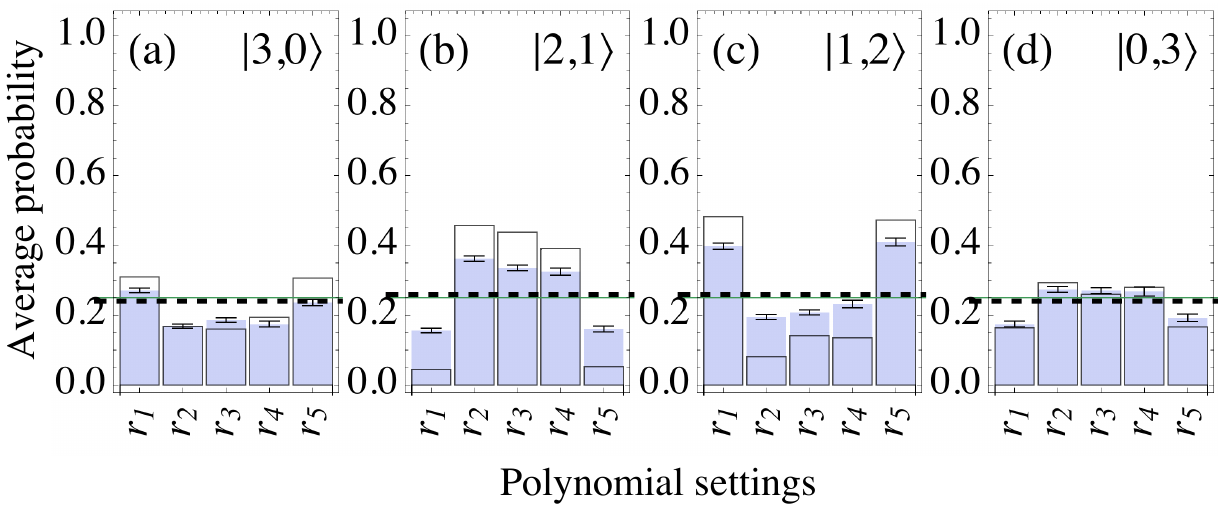}
\vspace{-0.7cm}
\caption{
\footnotesize{\textbf{Testing pseudorandomness with 3 photons.}
Detection outcomes 30, 21,  12, and 03 for three-photon interference in the two polarisation modes averaged over $\mathfrak{D}_2$. Solid colour represents measured data, while the empty boxes represent theory.  Error bars are computed from assuming Poisson-distributed noise on the photon detection statistics.  
The averages of $r_i$  over $\mathfrak{D}_2$ deviate sufficiently from the distinguishability corrected Haar value (green line), to indicate that $\mathfrak{D}_2$ cannot be a 3-design.
}}
\label{3photonAverage}
\end{figure}

To complete the characterisation of $\mathfrak{D}_2$, we use 3-photon interference for a set of five arbitrarily chosen physical degree-6 polynomial settings (labeled $r_1 , ... , r_5$).
These examples show that there exist polynomials whose average over $\mathfrak{D}_2$ differs from the Haar value.
The data is shown in Fig.~\ref{3photonAverage} and verifies that 
$\mathfrak{D}_2$ 
is not a 3-design.
Note that a single degree-6 polynomial that deviates from the Haar average is sufficient to show failure to simulate the Haar distribution. But due to the potential for increased indistinguishability---and therefore increased noise---in higher-photon number experiments using SPDC, we averaged five $r_i$ for good measure.

We quantify the similarity between the probability distributions extracted from our experiment and the ideal distributions predicted by a theoretical model by using the statistical fidelity between those and the discrete experimental distributions used to arrive at the averages shown in Fig.~\ref{1photonAverage}, \ref{2photonAverage} and \ref{3photonAverage}.
The average statistical fidelities are: $99.28 \pm 0.01 \%$ for the 300 2-photon experiments used for Fig.~\ref{2photonAverage} (a) and (b); $99.45\pm 0.02\%$ for the 300 2-photon experiments used for Fig.~\ref{2photonAverage} (c); and $97.55 \pm 0.03 \%$ for the 72 3-photon experiments used for Fig.~\ref{3photonAverage}.

Verification of pseudorandomness could also be conducted using process tomography~\cite{NielsenChuang} to estimate $x_1, y_1, x_2, y_2$, since then any desired polynomial could be computed and averaged over the ensemble to give the left-hand-side of (\ref{tdesigndefintion}).
However, this may not be possible in practice---for example in scenarios where probing individual elements of an ensemble is limited, such as a rapidly fluctuating quantum channel.
Consider a system
that produces unitary operations that are claimed to be drawn from the Haar-measure, or from a $t$-design with $t$ being sufficiently large for the random unitaries to be of value.
Standard tomography cannot characterise this scenario if we impose the restriction that each random output can only be probed with a single measurement on a single quantum state, since tomography requires repeated measurements on a fixed state in order to reconstruct quantum processes.

\begin{figure}[t!]
\includegraphics[width=0.9\columnwidth]{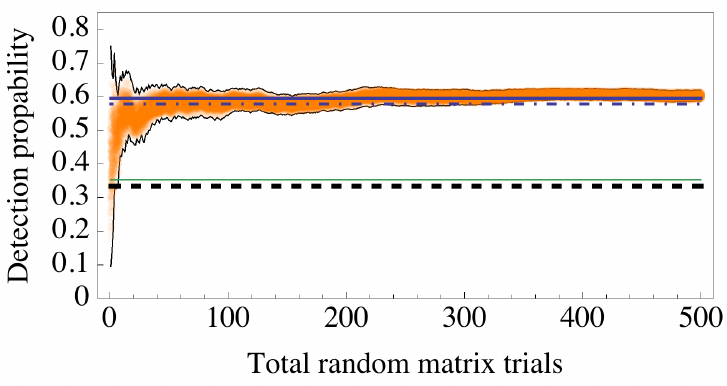}
\vspace{-0cm}
\caption{
\footnotesize{\textbf{Real-time failure of a photonic 1-design:}
Sixty-four independent estimates of the 2 photon probability distribution $|T_{1,1}T_{2,2}+T_{1,2}T_{2,1}|^2$ (overlaid orange points) for polynomial setting $q_{19}$, averaged over an increasing total number of random operations, each drawn uniformly from $\mathfrak{D}_1$.
}}
\label{ARF}
\end{figure}

An alternative approach to characterise randomness in this scenario 
is to average multiphoton interference over the randomised process. 
Fig.~\ref{ARF} shows the real-time failure of a photonic 1-design to behave Haar-randomly in the low photon rate regime. 
The wave plate configurations $U$ were set to realise uniformly at random one of the four $\mathfrak{D}_1$ operations, and $Q_\mathrm{in}$ and $Q_\mathrm{out}$ were fixed to realise the polynomial $q_{19}$.
For each implementation of $U$, we estimate the probability distribution of each 2-photon detection outcome using on the order of ten correlated detection events, which yields a noisy estimate of the distribution and by itself is insufficient to perform reliable process tomography. The total number of two-photon detection events for each random unitary lies in the range $[0,33]$, but this could in principle be performed with a quantum state for each $U$.
As we increase the number of samples of $\mathfrak{D}_1$ from 1 to 500, we compute a running average of $|{T}_{1,1}{T}_{2,2}+{T}_{1,2}{T}_{2,1}|^2$ which converges to a value of $0.603\pm0.001$.
For polynomial $q_{19}$, the theoretical value for a $1$-design should be 0.578 (marked with a blue dashed-dot line in the figure).
This process is repeated sixty-four times and overlaid (orange points) to observe statistical behaviour, while the upper and lower thin black lines bound the statistical error on computing average probability distributions from Poisson distributed noise on the photon detection rates.
The discrepancy between the experimental convergence and theory agrees with Fig.~\ref{2photonAverage} (b) and is attributed to imperfect wave plates and quantum interference.
Using characterised 2-photon interference only, we find that the experiment should converge to 0.594 (solid blue line). The black dashed line marks the Haar average detection probability, and it is clear within $\sim20$ trials that the random ensemble $\mathfrak{D}_1$ deviates significantly from this. 

We have realised a unitary 1-design and a unitary 2-design in two dimensions, using multiphoton interference for a complete verification of their pseudorandom properties. To do this, we have made use of the fact that multi-photon transition probabilities are polynomials in the matrix elements of the random process.
In general, $(t+1)$-photon states can be used to distinguish a $t$-design from a truly random ensemble of unitaries.
Furthermore, we have demonstrated a scenario where standard process tomography would be incapable of inferring the `degree' of randomness---given by a value of $t$---due to the lack of photon detection statistics collected for each individual unitary process, but where the accumulated average of correlated photon detection statistics is sufficient to show that a random process is not a unitary 2- (or higher) design.
We view these results as the first steps on a new path towards producing and studying pseudorandomness, an extremely valuable resource in quantum information technology.
For example, such ensembles will most likely be required to demonstrate conjectured extra-classical capabilities of linear quantum optics~\cite{ep-arxiv-1306.3995}.

\acknowledgements

We thank E. Hemsley, P. J. Shadbolt, X.-Q. Zhou, S. Bartlett, S. Vinjanampathy, H. Cable and J. Eisert for helpful discussions, {and especially T. Rudolph for the premise and advice}.
The authors are grateful for financial support from, EPSRC, ERC, NSQI. J.C.F.M. and P.S.T. are grateful for financial support from the EPSRC BGER. J.C.F.M. is supported by a Leverhulme Trust Early Career Fellowship. J.L.O.B. acknowledges a Royal Society Wolfson Merit Award and a RAE Chair in Emerging Technologies.

%\bibliography{JMatthewsTDoodles}

\newpage

\appendix

\section{Experimental setup details}
Here we provide details of the experimental setup depicted in Fig.~\ref{setup} (b).
The input photons for our experiment are generated from a pulsed parametric downconversion setup---a 808nm, 80MHz, fs pulsed Ti:sapphire laser is up-converted to 404nm light with a 2mm thick Bismuth Borate (BiBO) nonlinear crystal, the 404nm beam is then focused onto a BiBO crystal phase-matched for type-I parametric downcoversion.
We spatially select two paths that generate nominally identical photon pairs, polarised horizontally.
For the 1- and 2-photon experiments, we rotate the polarisation of path $b$ to vertical using a HWP, and combine the two paths of photon pairs onto one path using a polarising beam splitter cube.
This moves encoding to the two polarisation modes $H, V$ in one spatial mode $d$, and ideally all other degrees of freedom of the photons are made indistinguishable.
The photons then undergo quantum interference, after which the photons are detected at avalanche photodiodes, arranged in a number-resolving configuration using fibre splitters.
Recording data in the 2-photon coincidence basis, we post-select the two-photon term and zero loss.
The process is the same for 1-photon experiments, except the path $b$ in Fig.~\ref{setup} (b) is blocked and only 1-photon detection events are recorded.
To access 3-photon states, the polarisation in path $b$ is rotated to diagonal, and the 3-photon state $\ket{2}_H\ket{1}_V$ is post-selected as input into $T$ conditional on detecting 3 photons at the output and one photon in path c.
%The photon-number resolving setup of sixteen multiplexed detectors enables us to use post-selection to distinguish two-, four- and six-photon terms from the down-converted state.

\section{Wave plate settings}

The experiment configuration for the unitary rotations we investigate---highlighted by the dashed boxes in Fig.~1b of the main text---is modelled with $T~=~Q_\mathrm{out}~U~Q_\mathrm{in}$, where
%\ref{setup} b)---is modelled with $T=Q_\mathrm{out} U Q_\mathrm{in}$, where
\begin{eqnarray}
Q_\mathrm{in} & = &\mathcal{R}_{\textrm{QW\!P}}(\omega_2) \mathcal{R}_{\textrm{HW\!P}}(\omega_1)\label{bob1}\\
U &=& \mathcal{R}_{\textrm{QW\!P}}(\theta_3)\mathcal{R}_{\textrm{HW\!P}}(\theta_2) \mathcal{R}_{\textrm{QW\!P}}(\theta_1),\label{bob2}\\
Q_\mathrm{out} & = & \mathcal{R}_{\textrm{HW\!P}}(\omega_4) \mathcal{R}_{\textrm{QW\!P}}(\omega_3).\label{bob3}
\label{ExpConfig}
\end{eqnarray}
For the transition matrices for each half-wave plate (HWP) and quarter-wave plate (QWP) we take the convention that the angle of rotation is from the vertical:
\begin{eqnarray}
\mathcal{R}_{\textrm{HW\!P}}(\theta)=
\left(
\begin{array}{c c}
- i \cos{2\theta} & -i \sin{2\theta}\\
- i \sin{2\theta} & i \cos{2\theta}
\end{array}
\right), \label{hwpMat}\\
\mathcal{R}_{\textrm{QW\!P}}(\theta)=
\frac{1}{\sqrt{2}}\left(
\begin{array}{c c}
1- i \cos{2\theta} & -i \sin{2\theta}\\
- i \sin{2\theta} &1+ i \cos{2\theta}
\end{array}
\right). \label{qwpMat}
\end{eqnarray}
The settings used to configure $U$ for each element of the designs $\mathfrak{D}_1$ and $\mathfrak{D}_2$ (equations (3) and (4) of the main text) are given in Table~\ref{Usettings}. 
The wave plate rotation settings for the 9 independent degree-2 polynomials  $p_1,...,p_9$, the 25 independent degree-4 polynomials  $q_1,...,q_{25}$, and five arbitrarily chosen degree-6 polynomial  $r_1,...,r_5$ are given in Tables~\ref{psettings},~\ref{qsettings}~and~\ref{rsettings}. 
\begin{table}[h!]
\begin{center}
\begin{tabular}{c | c c c}
\hline
Pauli-operator notation &  $\theta_1$ & $\theta_2$ & $\theta_3$   \\ [0.5ex] % inserts table %heading
\hline
$I$ & 0 & 90 & 0 \\
$i X$ & 0 & -45 & 0\\
$-i Y$ & 45 & 90 & -45\\
$i Z$ & -45 & 90 & -45\\
$(I+i X-iY+iZ)/2$ & 0 & 90 & -45 \\
$(I+i X+iY+iZ)/2$ & -45 & 90 & 0 \\
$(I-i X-iY+iZ)/2$ & 45 & 90 & 0 \\
$(I-i X+iY+iZ)/2$ & 0 & 90 & 45 \\
$(I+i X-iY-iZ)/2$ & 45 & -45 & 0 \\
$(I+i X+iY-iZ)/2$ & 0 & -45 & 45  \\
$(I-i X-iY-iZ)/2$ & 0 & 45 & -45  \\
$(I-i X+iY-iZ)/2$ & -45 & 45 & 0  \\
\end{tabular}
\end{center}
\vspace{-0.5cm}
\caption{
\footnotesize{The wave plate settings for $\theta_i$ in degrees to realise the  elements of designs $\mathfrak{D}_1$ and $\mathfrak{D}_2$ (equation \ref{bob2}).
}}
\label{Usettings}
\end{table}
\begin{table}[h!]
\begin{center}
\begin{tabular}{c | c c c c || c | c c c c}
\hline
$p_i$ &  $\omega_1$ & $\omega_2$ & $\omega_3$ & $\omega_4$ & $p_i$ &  $\omega_1$ & $\omega_2$ & $\omega_3$ & $\omega_4$  \\ [0.5ex] 
\hline
$p_1$ & 0 & 0 & 0 & 0 & $p_6$ & 0 & 22.5 & 0 & 22.5 \\
$p_2$ & 0 & 0 & 0 & 22.5 & $p_7$ & 0 & 22.5 & 22.5 & 0 \\
$p_3$ & 0 & 0 & 0 & 45 & $p_8$ & 0 & 45 & 0 & 0 \\
$p_4$ & 0 & 0 & 22.5 & 0 & $p_9$ & 0 & 45 & 0 & 22.5 \\
$p_5$ & 0 & 22.5 & 0 & 0 & & & & &\\
\end{tabular}
\end{center}
\vspace{-0.5cm}
\caption{\footnotesize{Wave plate settings $\omega_i$ in degrees (equations \ref{bob1} and \ref{bob3}) for accessing a complete set of 9 independent degree-2 polynomials in 1-photon experiments to estimate $|T_{1,1}|^2$.
}}
\label{psettings}
\end{table}
\begin{table}[h!]
\begin{center}
\begin{tabular}{c | c c c c || c | c c c c}
\hline
$p_i$ &  $\omega_1$ & $\omega_2$ & $\omega_3$ & $\omega_4$ & $p_i$ &  $\omega_1$ & $\omega_2$ & $\omega_3$ & $\omega_4$  \\ [0.5ex] % inserts table %heading
\hline
$q_1$ & 0 & 0 & 0 & 0 & $q_{14}$ & 0 & 45 & 0 & 60 \\
$q_2$ & 0 & 0 & 0 & 22.5 & $q_{15}$ & 0 & 45 & 22.5 & 0 \\
$q_3$ & 0 & 0 & 0 & 60 & $q_{16}$ & 0 & 45 & 22.5 & 22.5 \\
$q_4$ & 0 & 0 & 22.5 & 0 & $q_{17}$ & 0 & 60 & 0 & 0 \\
$q_5$ & 0 & 0 & 22.5 & 22.5 & $q_{18}$ & 0 & 60 & 0 & 22.5 \\
$q_6$ & 0 & 0 & 45 & 22.5 & $q_{19}$ & 0 & 60 & 0 & 60 \\
$q_7$ & 0 & 22.5 & 0 & 0 & $q_{20}$ & 0 & 60 & 22.5 & 0 \\
$q_8$ & 0 & 22.5 & 0 & 22.5 & $q_{21}$ & 0 & 60 & 22.5 & 22.5 \\
$q_9$ & 0 & 22.5 & 0 & 60 & $q_{22}$ & 0 & 120 & 0 & 0 \\
$q_{10}$ & 0 & 22.5 & 22.5 & 0 & $q_{23}$ & 0 & 120 & 0 & 22.5 \\
$q_{11}$ & 0 & 22.5 & 22.5 & 22.5 & $q_{24}$ & 0 & 120 & 0 & 60 \\
$q_{12}$ & 0 & 45 & 0 & 0 & $q_{25}$ & 0 & 120 & 22.5 & 0 \\
$q_{13}$ & 0 & 45 & 0 & 22.5 & & & & & \\
\end{tabular}
\end{center}
\vspace{-0.5cm}
\caption{\footnotesize{Wave plate settings $\omega_i$ in degrees (equations \ref{bob1} and \ref{bob3}) for accessing a complete set of 25 independent degree-4 polynomials in 2-photon experiments to estimate $|T_{1,1}T_{2,2}+T_{1,2}T_{2,1}|^2$.
}}
\label{qsettings}
\end{table}
\begin{table}[h!]
\begin{center}
\begin{tabular}{c | c c c c || c | c c c c}
\hline
$p_i$ &  $\omega_1$ & $\omega_2$ & $\omega_3$ & $\omega_4$ & $p_i$ &  $\omega_1$ & $\omega_2$ & $\omega_3$ & $\omega_4$  \\ [0.5ex]
\hline
$r_1$ & 94.0   & 117.3   & 64.9   & 24.5  & $r_4$ & 179.7 & 11.36   & 24.6   & 108.1  \\
$r_2$ & 129.5 & 67.1    & 118.3  & 6.8  & $r_5$ & 1.9     & 114.0   & 162.5 & 160.7  \\
$r_3$ & 112.8 & 67.9    & 159.2  & 3.6  & 
\end{tabular}
\end{center}
\vspace{-0.5cm}
\caption{\footnotesize{Wave plate settings $\omega_i$ in degrees (equations \ref{bob1} and \ref{bob3}) to access a selection degree-6 polynomials using 3-photons.}}
\label{rsettings}
\end{table}

\newpage

\section{Basis monomials}
The unitaries $U$ we investigate are probed using transformations $Q_\mathrm{in}$ and $Q_\mathrm{out}$ to physically access polynomials in the elements of each $U$.
For 1-photon (2-photon) experiments the polynomials accessed are degree-2 (degree-4) in terms of the real variables $\{x_1, x_2, y_1, y_2\}$ that, together with the constraint $x_1^2~+~x_2^2~+~y_1^2~+~y_2^2~=~1$, parameterise SU(2) according to Equation (1) of the main text.
To prove linear independence (see \S D) of a set of polynomials, it is helpful to express them as a linear combination of basis monomials.
A complete and independent set of basis monomials
$m_i^2$
of degree-2 has 9 elements and is given by
\begin{eqnarray}
&&\left\{m^2_1, ... , m^2_9  \right\} \\
&&\,\,\,\,\,\,=\left\{x_1^2, x_1 x_2, x_2^2, x_1 y_1, x_1 y_2, x_2 y_1, x_2 y_2, y_1^2, y_1 y_2 \right\}\nonumber
\label{degree1monoms}
\end{eqnarray}
and a complete and independent set of basis monomials $m_i^4$ of degree-4 has 25 elements and is given by
\begin{eqnarray}
&&\left\{m^4_1, ... , m^4_{25}  \right\} \\
&&\,\,\,\,\,\,=\{x_1^4,
x_1^3 x_2,
x_1^2 x^2_2,
x_1 x_2^3,
x_2^4,
x_1^3 y_1,
x_1^2 x_2 y_1, \nonumber\\
&&\,\,\,\,\,\,\,\,\,\,\,\,x_1 x_2^2 y_1,
x_2^3 y_1,
x_1^3 y_2,
x_1^2 x_2 y_2,
x_1 x_2^2 y_2,
x_2^3 y_2,\nonumber\\
&&\,\,\,\,\,\,\,\,\,\,\,\,x_1^2 y_1^2,
x_1 x_2 y_1^2,
x_2^2 y_1^2,
x_1^2 y_1 y_2,
x_1 x_2 y_1 y_2,
x_2^2 y_1 y_2,\nonumber\\
&&\,\,\,\,\,\,\,\,\,\,\,\,x_1 y_1^3,
x_1 y_1^2 y_2,
x_2 y_1^3,
x_2 y_1^2 y_2,
y_1^4,
y_1^3 y_2\}\nonumber
\label{degree2monoms}
\end{eqnarray}

%\newpage

\section{Visual representation physical polynomial bases}

\begin{figure}[htp!]
\centering
\includegraphics[width=0.65\columnwidth]{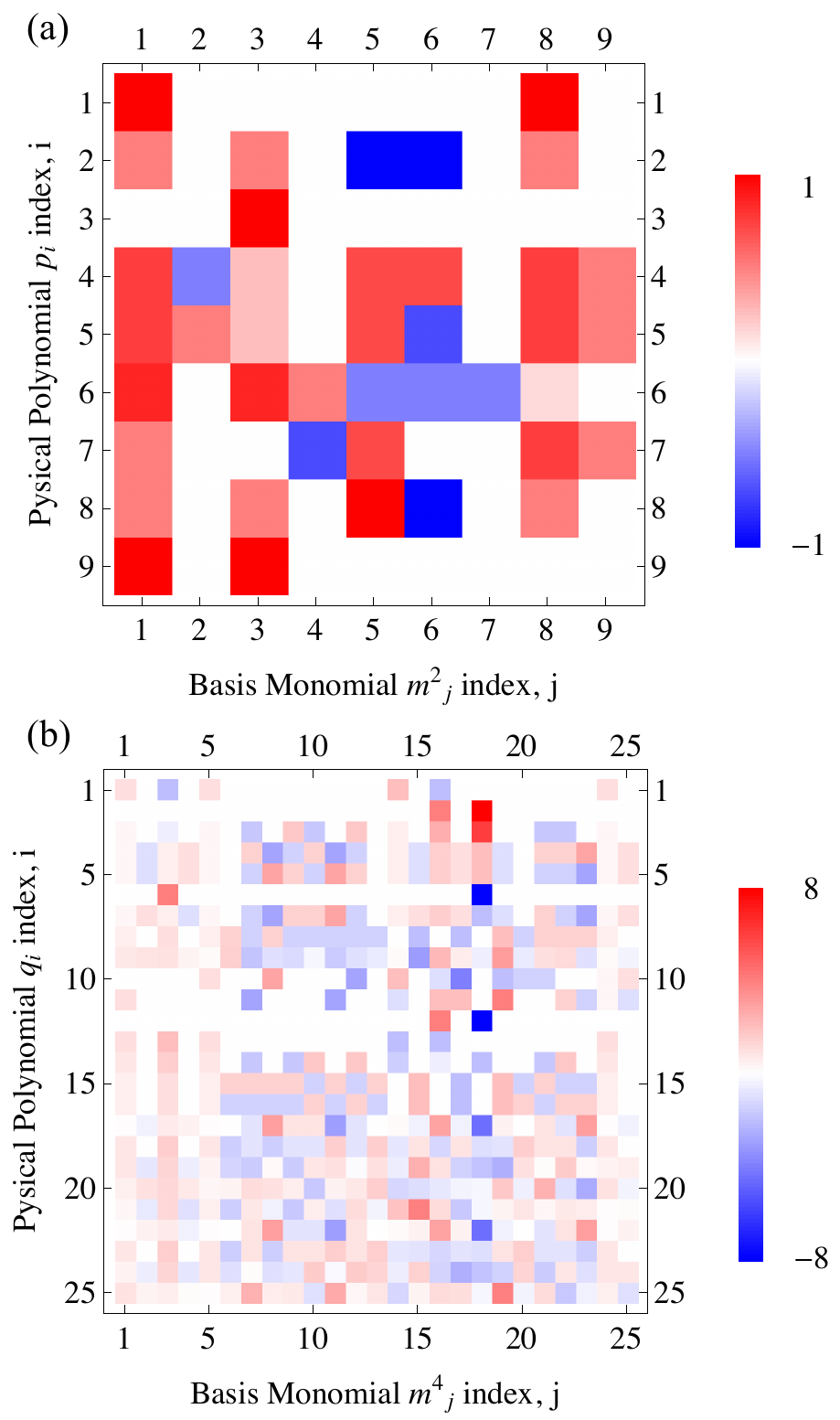}
%{PolyMatrix1Ph2Ph.pdf}
\caption{\footnotesize{Visualising (a) the matrix of physical polynomials $p_i$ and (b) the matrix of physical polynomials $q_i$ (see Appendix B), in terms of the basis monomials $m_j$ (see Appendix C).
}
}
\label{PolyMatrix1}
\end{figure}

To check that settings $p_i$ and $q_i$ (see Appendix B) provide access to a complete set of polynomials of degree-1 and 2 respectively, we substitute the corresponding wave plate settings into 
\begin{eqnarray}
T=\mathcal{R}_{\textrm{HW\!P}}(\omega_4)\mathcal{R}_{\textrm{QW\!P}}(\omega_3)U\mathcal{R}_{\textrm{QW\!P}}(\omega_2)\mathcal{R}_{\textrm{HW\!P}}(\omega_1),
\end{eqnarray}
where $U$ is parameterised by $x_1$, $x_2$, $y_1$, $y_2$ according to equation (1) of the main text.
We then construct a $9\times 9$ matrix where row $i$ corresponds to the polynomial $|T_{1,1}|^2$ for each $p_i$ setting and the $j$th element of that row corresponds to the coefficient of monomial $m^2_j$. 
It is then straightforward to confirm that this matrix has rank 9 and that the polynomials $|T_{1,1}|^2$ implemented with the $p_i$ settings form a linearly independent set. 
We follow the same procedure for the degree-4 polynomials, constructing a $25\times 25$ matrix that corresponds to $|T_{1,1}T_{2,2}+T_{2,1}T_{1,2}|^2$ for each $q_i$ setting and the coefficients of each monomial $m^4_j$. 
Again, it is then straightforward to confirm that this matrix has rank 25 and that the polynomials $|T_{1,1}T_{2,2}+T_{2,1}T_{1,2}|^2$ implemented with the $q_i$ settings form a linearly independent set. 
Both matrices are plotted in Fig.~\ref{PolyMatrix1}.

\section{Analysis of uniformity}
Figures {2}, {3} and {4} of the main text compare a series of polynomial settings, averaged over finite ensembles of unitary operations, to the expected Haar average for each case. 
Since these polynomials are normalised probabilities, the expected Haar average will be constant over all polynomial settings. 
\begin{figure}[b!]
\includegraphics[width=1\columnwidth]{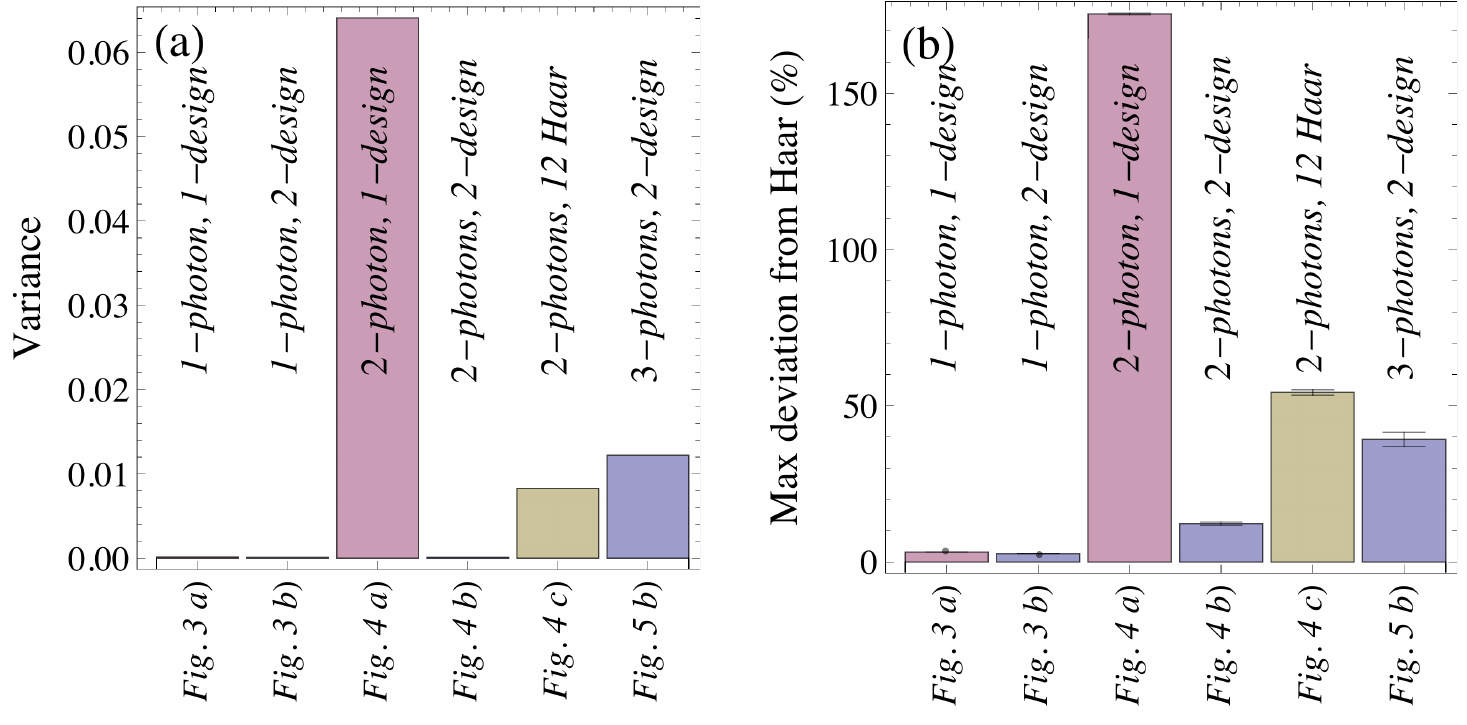}
%{Uniformity.pdf}
\caption{\footnotesize{
\textbf{Uniformity of averages over finite ensembles.} Uniformity of each data set of the main text, as labeled, is plotted in terms of (a) the variance over the set of polynomials measured and (b) the maximum deviation of the average probability, as a percentage, from the expected Haar average---black dashed line in Fig.~{2} of the main text and green solid lines distinguishability corrected Haar value (see Appendix G) in Figs.~{3} and {4} of the main text.
}}
\label{Uniformity}
\end{figure}

How accurately the finite ensembles we realise mimic Haar distributed unitary matrices can be quantified by the uniformity of the average probabilities over the different polynomials settings, and how much they deviate from the expected Haar average in each case. 
We quantify the uniformity of the ensemble behaviour plotted in Figs.~{2}, {3} and {4} of the main text in Fig.~\ref{Uniformity}.
Note that while theory predicts that $t$-photon interference over a $t$-design should yield perfectly uniform results, experimental imperfections in realising design elements can give rise to non-uniformity, as we observe here. 
A topic of future work will be to quantify the effect of experimental imperfections on simulating Haar randomness.

\section{Fidelities of the experimentally extracted probability distributions}

Experimental error arise due to imperfect optical components, imperfect control of those components and non-unit quantum interference. 
For the case of 3-photon interference, we suffer temporal distinguishability that degrades the quality of quantum interference from multiphoton terms generated within the same pulse of the laser pumping the parametric down conversion process.

\begin{figure}[h!]
\includegraphics[width=0.85\columnwidth]{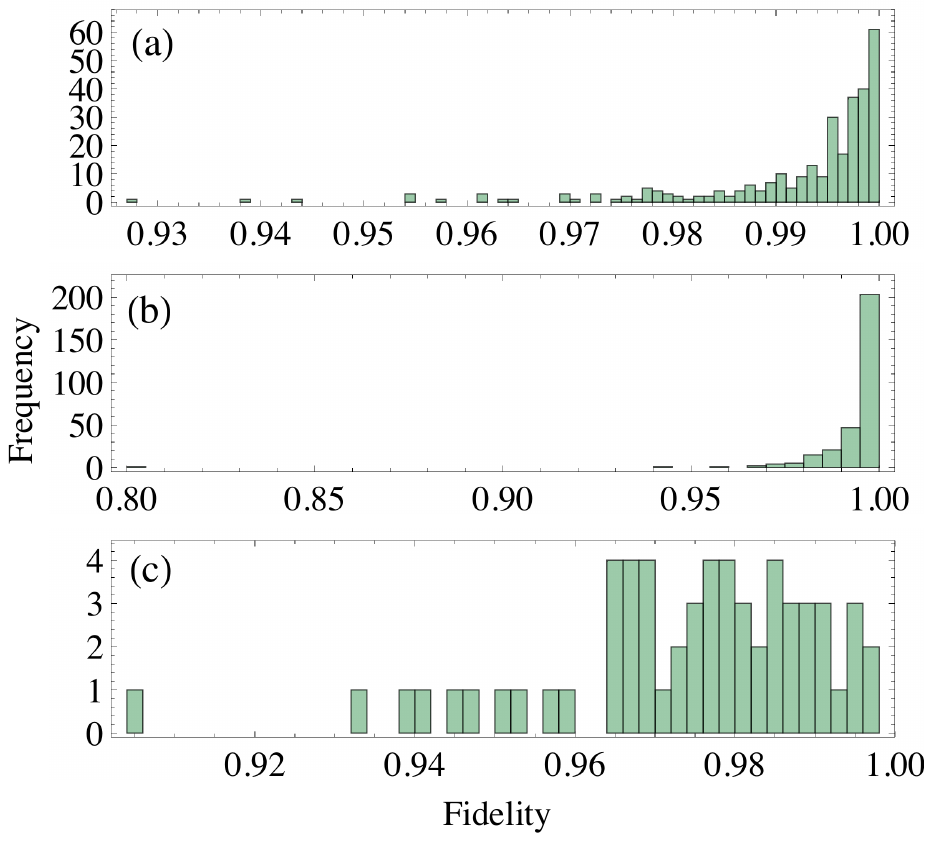}
%{FildelitiesHis2&3Pho200913.pdf}
\caption{\footnotesize{
\textbf{Comparing experiment with theory.} Histograms of fidelities between theoretically predicted distributions with the corresponding distributions extracted from experiment for: (a) 2-photon interference in each element of $\mathfrak{D}_2$, for all 25 linearly independent polynomial settings $\{q_1,...,q_{25}\}$ (Fig.~\ref{2photonAverage} (a,b) of the main text and Fig.~\ref{2PhotonFull} (a - f)); (b)  2-photon interference in 12 Haar-randomly chosen processes, for all 25 linearly independent polynomial settings  $\{q_1,...,q_{25}\}$ (Fig.~\ref{2photonAverage} (c) of the main text and Fig.~\ref{2PhotonFull} (g - i)) and (c) 3-photon interference in each element of $\mathfrak{D}_2$ for 5 randomly chosen linearly independent polynomial settings $\{r_1,...,r_5\}$ (Fig.~{4} (a - d) of the main text).
}}
\label{Fids}
\end{figure}

To quantify the quality of the probability distributions extracted from our experiments, we use the fidelity between two discrete probability distributions $P$ and $P'$ given by
\begin{eqnarray}
F(P,P') = \sum_{i}\sqrt{P_i P'_i}
\end{eqnarray}
We show in Fig.~\ref{Fids} histograms of the fidelities of the 300 two-photon experiments using 
$\mathfrak{D}_2$,
the 300 two-photon experiments using 12 Haar-randomly chosen SU(2) elements and the 72 three-photon experiments.

\section{{Mode mismatch characterisation with HOM interference}}

In Fig. {2}, {3} and {4} of the main text and Fig.~\ref{ARFFull} and \ref{2PhotonFull}, we plot as black dashed lines the ideal 2- and 3-photon interference distributions averaged over the Haar measure for each case. 
This value is altered by imperfections in the distinguishability of photons. 
However, photon distinguishability does not alter the uniformity of the  average degree-$2t$ polynomials over a $t$-design or the Haar measure.

For example, in the two-photon interference displayed in Fig.3, distinguishability can be modelled with a single parameter $\theta$ which alters the two-photon detection statistics of the Hong-Ou-Mandel experiment in unitary operation $T$ according to
\begin{eqnarray}
&&\cos^2{\theta}
\left|
T_{1,1}T_{2,2} + T_{1,2} T_{2,1}
\right|^2
\nonumber \\
&&+ \sin^2{\theta}\left(
\left|
T_{1,1}T_{2,2}\right|^2
+
\left|T_{1,2} T_{2,1}\right|^2
\right)
\end{eqnarray}
Similarly, the mode-mismatch parameter alters the distribution of three-photon interference. 
Since distinguishability merely changes the form of the polynomial, the $t$-photon interference distributions averaged over $t' \geq t$-designs and the Haar measure is still uniform.

\begin{figure}[b]
\centering
\includegraphics[width=0.9\columnwidth]{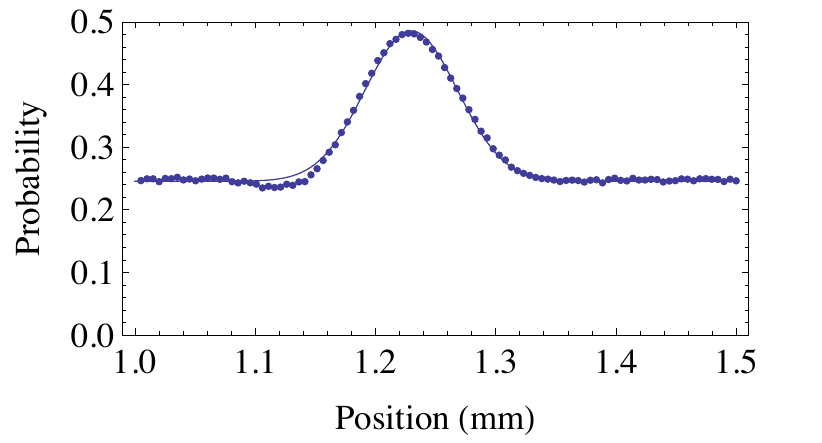}
%{FittedDip200913.pdf}
\caption{\footnotesize{
The probability for two photons to bunch at one output in a Hong-Ou-Mandel experiment, as interfering modes overlap temporally. 
We observe a visibility is $V=-0.941724$ using a Gaussian best fit.
}}
\label{HOM}
\end{figure}

We characterise the mode mismatch parameter $\theta$ for modes $H$ and $V$ in path $d$ (see Fig.~{1} of the main text) using two photon interference in the wave plate configuration $T = Q_{out} U Q_{in}$. 
Irrespective of the unitary, the probability to detect two photons in the same output should go from $1/4$ in the distinguishable case to $1/2$ when both photons are overlapped. 
We control this overlap via optical delay $\Delta \tau$ shown in Fig. 1b). 
Ideally this produces a ``peak'' in interference with visibility of $V = -1$. 
Any mode mismatch, i.e. $\theta \neq 0$ degrades this interference. 
In our setup, we measure $V = -0.942$ (Fig.~\ref{HOM}), which we then use to compute the mode mismatch parameter: $\theta=0.244$ radians.

Using this value of $\theta$, we compute the Haar average of the 2- and 3-photon interference distributions in the presence of $\theta=0.244$ distinguishability between the $H$ and $V$ modes of our experiment. 
This adjusted Haar average is given by the green lines of Figs.~{3}, {4}, and {5} of the main text and Figs.~\ref{ARFFull} and \ref{2PhotonFull}.

Also the solid blue line in Fig.~{5} marks the expected average of 2-photon interference for $\theta=0.244$, averaged over the 1-design $\mathfrak{D}_1$ and this mode-mismatch parameter is taken into account to compute the theoretical distributions of each individual experiment; these are represented by the empty bar charts in Figs.~{3} and {4} of the main text and Figs.~\ref{ARFFull} and \ref{2PhotonFull}.

\section{Loss characterisation}

\begin{figure}[htp!]
\centering
\includegraphics[width=0.5\columnwidth]{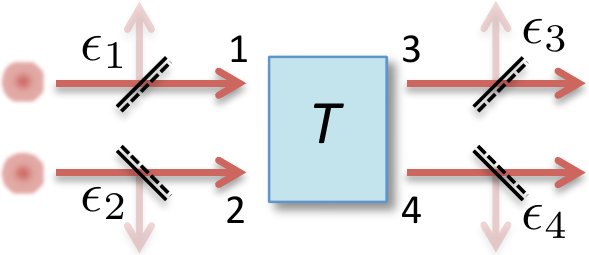}
%{LossCharacterisation.pdf}
\caption{\footnotesize{
Loss channels at the input and output of unitary processes can be modelled as beamsplitters with reflectivity $\epsilon_i$.
}}
\label{LossFig}
\end{figure}

The losses present in our system are illustrated in Fig.~\ref{LossFig}. 
On inputting a photon pair into inputs 1 and 2, the input loss rates $\epsilon_1$, $\epsilon_2$ effect the state input into $T$ according to
\begin{eqnarray}
a^\dagger _1 a^\dagger _2 \stackrel{\textrm{Input loss}}{\longrightarrow} \sqrt{\epsilon_1 \epsilon_2} a^\dagger _1 a^\dagger _2
\end{eqnarray}
Post-selection of detecting two photons at the experiment enable us to ignore the effect of $\epsilon_1$ and $\epsilon_2$ and observe the effects of $T$ acting on our input state. 
Ignoring the input losses is possible for any input state of the form ${a^\dagger _1}^n {a^\dagger_2}^m$ when post-selecting a total of $m+n$ photons at the output.

Unitary evolution maps a symmetric input state to a symmetric output state. 
This fact can be used to cancel the effects of output loss. 
Ignoring $\epsilon_1$ and $\epsilon_2$, a photon pair input into 1 and 2 evolves in Fig.~\ref{LossFig} according to
\begin{eqnarray}
{\epsilon_3}T_{1,1} T_{2,1}  a^\dagger _3 a^\dagger _3+ \sqrt{{\epsilon_3}{\epsilon_4}}(T_{1,1} T_{2,2} +T_{2,1} T_{1,2}) a^\dagger_3 a^\dagger_4\nonumber\\
+ {\epsilon_4}T_{1,2} T_{2,2} a^\dagger_4 a^\dagger_4
\end{eqnarray}
where we have used the elements of the matrix representation of $T$. 
Since $T\in SU(2)$, we know from equation (1) of the main text that $|T_{1,1} T_{2,1}|^2 = |T_{1,2} T_{2,2}|^2$. 
In the lossless case this means the probability to detect two photons at output 3 is equal to the probability to detect two photons at output 4; $P(2,0) = P(0,2)$.

The total number of photons, $C_{\textrm{total}}$, entering $T$ are distributed across to the detection rates
\begin{eqnarray}
C(2,0) &=& \epsilon_3^2 P(2,0) C_{\textrm{total}}\nonumber\\
C(1,1) &=& \epsilon_3 \epsilon_4 P(1,1) C_{\textrm{total}}\\
C(0,2) &=& \epsilon_4^2 P(0,2) C_{\textrm{total}}\nonumber
\end{eqnarray}
Since we are using post selection here, it is sufficient to determine the relative loss between channel 3 and 4. 
We can therefore assume $\epsilon_4=1$; from this it follows
\begin{eqnarray}
\epsilon_3 = \sqrt{C(2,0)/C(0,2)}
\end{eqnarray}
Since this is computed from photon coincidence count rates, Poisson distributed noise is assumed to be present on $C(2,0)$, $C(1,1)$ and $C(0,2)$, which propagates into the characterisation of $\epsilon_3$ and the subsequently computed normalised probability distributions presented.

\section{Full multiphoton probability distributions}

\begin{figure}[h!]
\centering
\includegraphics[width=0.8\columnwidth]{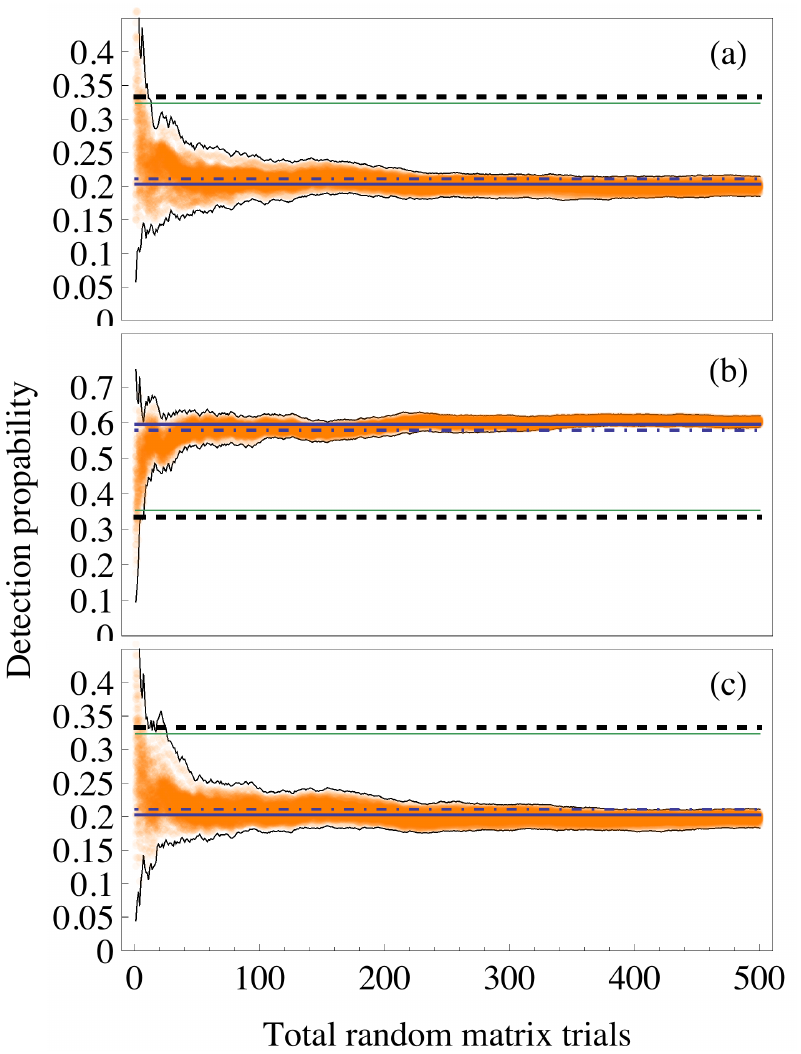}
%{ARFapp200913.pdf}
\caption{\footnotesize{
All 2-photon probabilities corresponding to Fig.~{5} in the main text.
}}
\label{ARFFull}
\end{figure}

\begin{figure*}[t!]
\centering
\includegraphics[width=1.8\columnwidth]{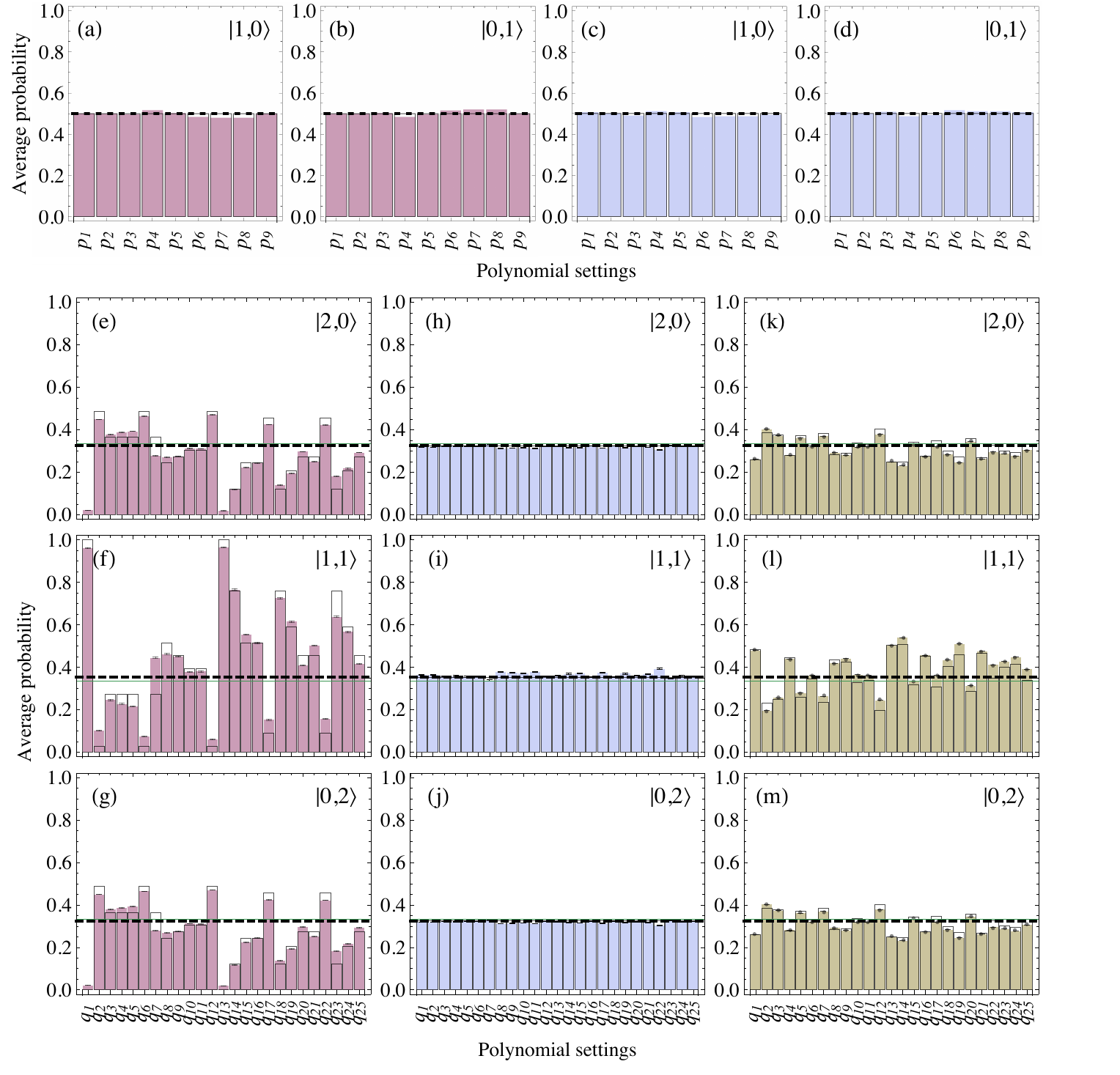}
%{AppendixAllProbs.pdf}
\caption{\footnotesize{
All 1-photon (a-d) and 2-photon (e-m) probabilities corresponding to Figs.~{2} and Fig.~{3} of the main text.
}}
\label{2PhotonFull}
\end{figure*}

For completeness, Fig.~\ref{ARFFull} and Fig.~\ref{2PhotonFull} display the full probability distributions of photon detection events extracted from experiment.

\end{document}